\documentstyle[12pt]{article}
\renewcommand \baselinestretch{1.4}

\begin{document}

\def\beq{\begin{equation}}
\def\eeq{\end{equation}}
\def\bce{\begin{center}}
\def\ece{\end{center}}
\def\bea{\begin{eqnarray}}
\def\eea{\end{eqnarray}}
\def\ben{\begin{enumerate}}
\def\een{\end{enumerate}}
\def\ul{\underline}
\def\ni{\noindent}
\def\nn{\nonumber}
\def\bs{\bigskip}
\def\ms{\medskip}
\def\wt{\widetilde}
\def\wh{\widehat}
\def\tr{\mbox{Tr}\, }
\def\brr{\begin{array}}
\def\err{\end{array}}



\hfill June, 1995

\vspace*{3mm}

\begin{center}

{\LARGE \bf
 Dilatonic gravity  near two dimensions as a string theory}

\vspace{8mm}

\renewcommand
\baselinestretch{0.5}
\medskip

{\sc E. Elizalde}
\footnote{Permanent address:
Center for Advanced Study CEAB, CSIC, Cam\'{\i} de Santa
B\`arbara, 17300 Blanes,
Spain;
e-mail: eli@zeta.ecm.ub.es} \\
Dipartimento di Fisica, Universit\`a degli Studi di Trento, \\
I-38050 Povo, Trento, Italia \\
 and \\
{\sc S.D. Odintsov} \footnote{On
leave from: Tomsk Pedagogical Institute, 634041 Tomsk, Russia;
e-mail: odintsov@ecm.ub.es}
\\
Department ECM, Faculty of Physics,
University of  Barcelona, \\  Diagonal 647, E-08028 Barcelona,
Spain \\

\vspace{15mm}

{\bf Abstract}

\end{center}

Using the renormalization-group formalism, a sigma model of a
special
type ---in which the metric and the dilaton depend explicitly on one
of
the string coordinates only--- is investigated near two dimensions.
It is seen that
dilatonic gravity coupled to $N$ scalar fields can be expressed in
this form, using a string parametrization, and that it possesses the usual
UV
fixed point. However, in this stringy parametrization of the
theory
the fixed point for the scalar-dilaton coupling turns out to be trivial,
while the
fixed point for the gravitational coupling remains the same as
in previous studies being, in particular, non-trivial.

\vspace{4mm}


\newpage

\section{Introduction}

Having its origin in work by Weinberg \cite{4}, where he proposed a
program that should lead to the realization of an asymptotically
safe
quantum gravity (QG), a strong belief exists nowadays that a
consistent theory of QG might be actually constructed by means of
a
process of analytical continuation from two to four dimensions. Further
work
in this direction has shown however (see \cite{4} and references
therein), that $(2+\epsilon)$-dimensional Einsteinian gravity ought
 to be
probably ruled out from this program, as a consequence of its
non-renormalizability at the two-loop level (see the paper by Jack
and
Jones in Ref. \cite{4}). Recently, there has been the suggestion
\cite{5}-\cite{9}
to study dilatonic gravity (with matter) in $2+\epsilon$
dimensions.
This theory has a smooth behavior in the limit $\epsilon
\rightarrow 0$  and is renormalizable near two dimensions.
The
existence of a non-trivial ultraviolet fixed point (a saddle point
actually) has been observed in Refs. \cite{5}-\cite{9} for different
versions
of dilatonic gravity with matter near two dimensions. However, in
dilatonic gravity the beta-functions depend explicitly on the
background
dilaton field, as in string theory \cite{1,2}. Both are defined
through the use of the standard off-shell effective action and, as
a
result, they are dependent on the gauge and also on the field
parametrization. Hence, the position of the fixed point in
dilatonic
gravity is gauge dependent too \cite{6}. It is interesting to study
this
issue further, and to try to ascertain if dilatonic gravity is
really
an asymptotically safe theory, e.g. of the kind mentioned above.

In this letter we shall study  dilatonic
gravity with scalar matter
near two dimensions. We shall use a parametrization of a special
kind
(a stringy parametrization),
in which dilatonic gravity is represented under the form of the
standard sigma model
\cite{10,11}. Then, the renormalization of this sigma model near
two
dimensions will be discussed and the corresponding beta-functions
will
be obtained. By studying dilatonic gravity in $2+\epsilon$
dimensions in
this stringy parametrization we will find that an UV non-trivial
fixed
point for the gravitational coupling constant appears which is the
same
as the one obtained in Refs. \cite{5}-\cite{9}. However, the fixed point
for
the dilaton-scalar coupling will turn out to be a trivial one,
contrary
to the situation that has been observed in Refs. \cite{5}-\cite{9}. To
finish, we
will briefly comment on this original new result.

\section{A sigma-model of a special type in $2+\epsilon$
dimensions}

Our starting point will be the standard $\sigma$-model action
corresponding to string theory \cite{10,11}
 \beq
S = \int d^d x\,  \sqrt{-g} \left[{1 \over 2} G_{ij} (X)
g^{\mu\nu}
\partial_\mu X^i \partial_\nu X^j + {1 \over 16\pi
G} R \phi (X) \right], \label{1}
\eeq
where $i,j = 1, \ldots, D$,
 $g_{\mu\nu} (x) $ ($\mu,\nu =1, \ldots, d$, $d=2+\epsilon$) is the
($2+\epsilon$)-dimensional
metric, $R$ the corresponding curvature, and $\phi (X)$ is the
dilaton.
In the standard sigma-model approach to string theory it is $d=2$,
and
in the conformal gauge the one-loop effective action is given by
 \beq
\Gamma_{div} = \frac{1}{2\pi \epsilon} \int d^2 x\,  \sqrt{-g}
\left[{1 \over 2} \beta_{ij} (X)
g^{\mu\nu} \partial_\mu X^i \partial_\nu X^j +  \beta_\phi (X) R
\right], \label{2}
\eeq
where \cite{1,2}
\beq
\beta_{ij} (X)= R_{ij}, \qquad   \beta_\phi (X)  = \frac{26 -D}{12}
- 8\pi \frac{1}{G} \, \Box^{(D)} \phi (X).
\label{2c}
\eeq
Notice that the beta-functions of the above sigma model (which are
{\it not} the RG $\beta$-functions in our discussion) are generically
different from the Weyl anomaly coefficients \cite{3}, and also,
that the loop-expansion parameter in our analysis is $G$.

We now use the fact that the theory (\ref{1}) for $d=2+ \epsilon$
has a smooth limit for $\epsilon \rightarrow 0$. As has been
shown, there are no problems when $\epsilon \rightarrow 0$ in this
theory, unlike what happens in the case of Einstein's gravity near
two dimensions \cite{4}. In other words, one can use the results of
the calculation of the effective action at exactly-two
dimensions, in order to obtain the one-loop effective action near
two dimensions, as it was the case with dilatonic gravity
\cite{5,6}. Then, the renormalization of the theory can be carried
out in close analogy with the case of dilatonic gravity \cite{5,6}.

For simplicity we will consider the situation where the sigma-model
under study is such that $G_{ij}$ and $\Gamma_{div}$ (i.e.,
$\beta_\phi$ and $\beta_{ij}$) depend explicitly on a single field
$\varphi \in \left\{ X^i \right\}$ {\it only}. The dilaton in
(\ref{1}) is also conveniently parametrized as $\phi (X)
\rightarrow e^{-2\phi (X)}$ and is to be considered as depending only on
the field $\varphi$, so that in the end it is always possible
to write $\phi (X) = e^{-2\varphi}$.

The renormalization of the fields can be done as in dilatonic
gravity \cite{5}, that is
\bea
g_{0\mu\nu} = g_{\mu\nu}e^{-2\Lambda (\varphi )} && (\Lambda
(0)=0), \nn \\ \varphi_0 = \varphi  + f(\varphi) && (f(0)=0),
\nn \\ G_{0ij} = G_{il} (\varphi) Z_{lj} (\varphi ), && \label{3}
\eea
where $Z_{lj} = \delta_{lj} + \wt{Z}_{lj}$. In this case the bare
action $S_0$ is given by
\bea
S_0 &=& \int d^d x\,  \sqrt{-g} \left\{ {1 \over 16 \pi G_0} R e^{-
2\varphi - \epsilon \Lambda -2 f} \right. \nn \\ &&+ \frac{\epsilon
+1}{16\pi G_0} \left[ 4\Lambda' + \epsilon (\Lambda')^2 +
4f'\Lambda' \right]  e^{-2\varphi - \epsilon \Lambda -2 f}
g^{\mu\nu} \partial_\mu \varphi \partial_\nu \varphi \nn \\ && +
\left. \frac{1}{2} G_{il} Z_{lj} (\varphi) e^{-\epsilon \Lambda}
g^{\mu\nu} \partial_\mu X^i \partial_\nu X^j  \right\}, \label{4}
\eea
where $\Lambda' = \partial \Lambda /\partial \varphi$. One has to
compare (\ref{4}) with $S_0=S + S_{counter}$, where $S_{counter} =
- \mu^\epsilon \Gamma_{div}$. By just matching corresponding terms,
we get
\bea
&& {1 \over 16 \pi G_0}  e^{-2\varphi - \epsilon \Lambda -2 f} =
\mu^\epsilon \left(  {1 \over 16 \pi G}  e^{-2\varphi } - {1 \over
2 \pi \epsilon} \beta_\phi (\varphi) \right), \nn \\
&&  \frac{\epsilon +1}{16\pi G_0} \left[ 4\Lambda' + \epsilon
(\Lambda')^2 + 4f'\Lambda' \right]  e^{-2\varphi - \epsilon \Lambda
-2 f} + \frac{1}{2} G_{\varphi l} Z_{l\varphi} (\varphi) e^{-
\epsilon \Lambda} = \frac{1}{2} G_{\varphi \varphi} -
\frac{\mu^\epsilon}{4\pi \epsilon} \beta_{\varphi \varphi}, \nn \\
&&  G_{il} Z_{lj} (\varphi) e^{-\epsilon \Lambda}  = G_{ij} -
\frac{\mu^\epsilon }{2\pi \epsilon} \beta_{ij},
 \label{5}
\eea
where in the last expression the case $i=j=\varphi$ is excluded and
the index $i$ corresponding to $\varphi \in \left\{ X^i \right\}$
is also denoted as $\varphi$.

We will now take into account the fact that $G$ is the loop
expansion parameter, and that $f,\Lambda \sim {\cal O} (G)$.
Dropping terms of higher order in $G$ and following the procedure
of Ref. \cite{5}, Eqs. (\ref{5}) can be written as
\bea
&& {1 \over 16 \pi G_0}  e^{-2\varphi - \epsilon \Lambda (\varphi)
-2 f (\varphi)} = \mu^\epsilon \left(  {1 \over 16 \pi G}  e^{-
2\varphi } - {1 \over 2 \pi \epsilon} \beta_\phi (\varphi) \right),
\nn \\
&&  \frac{\epsilon +1}{4\pi G_0} \Lambda' (\varphi)   e^{-2\varphi}
+ \frac{1}{2} G_{\varphi l} Z_{l\varphi} (\varphi) e^{-\epsilon
\Lambda (\varphi)} = \frac{1}{2} G_{\varphi \varphi} -
\frac{\mu^\epsilon}{4\pi \epsilon} \beta_{\varphi \varphi}
(\varphi), \nn \\ &&  G_{il} Z_{lj} (\varphi) e^{-\epsilon \Lambda
(\varphi)}  = G_{ij} - \frac{\mu^\epsilon }{2\pi \epsilon}
\beta_{ij} (\varphi),
 \label{6}
\eea
where expansion of the exponential functions up to linear terms in
$G$ is to be understood. From Eqs. (\ref{6}) one can get explicit
relations between the non-renormalized and the renormalized
parameters. In particular,
\beq
\frac{1}{G_0} = \mu^\epsilon \left( \frac{1}{G} -
\frac{8}{\epsilon} \beta_\phi (0) \right).
\label{7}
\eeq
For the rest of the parameters of the theory the renormalization
can be performed in different ways (that turn out to be ambiguous).
Hence, from now on we will restrict our considerations to an even
smaller class of sigma models, where $G_{\varphi \varphi} -
G_{\varphi l} Z_{l\varphi} =0$ and $\beta_{ij} =0$ (notice that
this case is consistent with the choice of $G$ as the loop
expansion parameter).  Then $\Lambda (\varphi) $ can be easily
found, from the second of Eqs. (\ref{6}), to be:
\beq
\Lambda (\varphi) = - \frac{G}{\epsilon +1} \int_0^\varphi
e^{2\varphi'} \frac{\beta_{\varphi \varphi} (\varphi')}{\epsilon}
\, d\varphi'.
\label{8}
\eeq
Substituting (\ref{7}) and (\ref{8}) into the first and into the last
of Eqs. (\ref{6}) one finds, to leading order in $G$,
\bea
&&f(\varphi)= \frac{4G\beta_\phi (0)}{\epsilon} \left(e^{2\varphi}
-1 \right) +  \frac{G}{2(\epsilon +1)} \int_0^\varphi e^{2\varphi'}
\beta_{\varphi \varphi} (\varphi') \, d\varphi'
+ \frac{4G}{\epsilon}e^{2\varphi} \left[ \beta_\phi (\varphi)-
\beta_\phi (0)\right], \nn \\
&& \wt{Z}^k_{\ j} (\varphi)= - \frac{\delta^k_{\ j}G}{\epsilon +1}
\int_0^\varphi e^{2\varphi'} \beta_{\varphi \varphi} (\varphi') \,
d\varphi'.
\label{9}
\eea

We can now turn to the evaluation of the beta functions of the
sigma model under discussion. We get for $G$
\beq
\beta_G = \mu {\partial G \over \partial \mu} = \epsilon G - 8
 G^2 \beta_\phi (0).
\label{10}
\eeq
This beta function comes from the renormalization of the dilaton in
the original string theory.
In the calculation of the beta-function for the metric we will be
interested in its dilatonic (i.e. $\varphi$-) dependence only. In
this case it is enough to consider the renormalization of  $G_{ij}$
which depends on $\varphi_0$ (to leading order in $G$):
\bea
G_{0ij} (\varphi_0) &=& G_{ij} (\varphi) -
 \frac{G_{ij}(\varphi) G }{\epsilon +1} \int_0^\varphi
e^{2\varphi'} \beta_{\varphi \varphi} (\varphi') \, d\varphi' \nn
\\
&=& G_{ij} (\varphi_0) - f (\varphi_0) G_{ij}' (\varphi_0)
- \frac{G_{ij}(\varphi) G }{\epsilon +1} \int_0^\varphi
e^{2\varphi'} \beta_{\varphi \varphi} (\varphi') \, d\varphi'.
\label{11}
\eea
Then,
\beq
\beta_{G_{ij}}  =\mu {\partial G_{ij} (\varphi_0) \over
\partial \mu} =\epsilon f (\varphi_0) G_{ij}' (\varphi_0)
+ \frac{\epsilon G_{ij}(\varphi) G }{\epsilon +1}
\int_0^\varphi e^{2\varphi'} \beta_{\varphi \varphi} (\varphi') \,
d\varphi'.
\label{12}
\eeq
With this beta function we terminate here the construction of the string RG
near two dimensions. We will turn now to consider a physically
interesting example.

\section{Dilatonic gravity near two dimensions as a string theory}

Dilatonic gravity provides an interesting example of a string
theory of the above type near two dimensions. The corresponding
Lagrangian can be written as
\beq
L=C(\varphi ) R - \frac{1}{2} e^{-2\phi (\varphi )} \partial_\mu
\chi_a \partial_\nu \chi^a,
\label{13}
\eeq
where $\varphi$ is the dilaton and $\chi_a$ are $N$ scalar fields.
This is a popular toy model for the study of quantum gravity and
black hole physics. Now, using the method of Ref. \cite{7} in the
conformal gauge
\beq
g_{\mu\nu} = e^{2\sigma} \bar{g}_{\mu\nu},
\label{14}
\eeq
one can represent the action of dilatonic gravity as a sigma model,
in the form
 \beq
S = \int d^2 x\,  \sqrt{-g} \left[{1 \over 2} G_{ij} (X)
\bar{g}^{\mu\nu}
\partial_\mu X^i \partial_\nu X^j +  R \phi (X) \right], \label{15}
\eeq
where
\beq
G_{ij} = \left( \brr{cccc} 0 & 2C'(\varphi ) & | & 0 \\
 2C'(\varphi ) &0 &  | & 0 \\ --- & --- & | & --- \\
 0 & 0 & | & -e^{-2\phi (\varphi)} \err \right), \qquad X^i = \left\{
\varphi, \sigma, \chi_a \right\}, \qquad \phi (X) = C(\varphi).
\eeq
Using Weyl rescaling of the two-dimensional metric one can
parametrize the initial Lagrangian (\ref{13}) so that $C(\varphi)=
e^{-2\varphi} /(16\pi G)$.

One can easily realize that the sigma model (\ref{15}) is of the same
type as the one described in the previous section.
Turning to the construction of the geometry of such a sigma model,
we obtain the stringy $\beta$-functions:
\bea
\beta_\phi (\varphi ) &\equiv & \beta_\phi (0) = \frac{24 -N}{12},
\nn \\
\beta_{\varphi\varphi} &=& R_{\varphi\varphi}=N \left[ (\phi')^2
(\varphi) - \phi'' (\varphi) - 2\phi'(\varphi) \right],
\label{16}
\eea
and $\beta_{ij} =0$ for the rest of the indexs $i,j$. Notice that
$\Gamma_{div}$ in this model has been calculated also in Refs.
\cite{8}, \cite{6} and \cite{5}, in different gauges. The results
in the three references are all different from one another, and
also different from Eq. (\ref{16}) (note, however, the coincidence
of (\ref{16}) with the result in Ref. \cite{6} in the gauge $\alpha
=0$). This should not be considered as strange, taking into account
the fact that
$\Gamma_{div}$ is gauge dependent off-shell and that the effective
action is also parametrization dependent. Once we use the classical
equations of motion, all the calculations (\cite{8,6,5} and
(\ref{16})) lead to the same effective action on shell, what can be
checked easily (see the Appendix).

Substituting  (\ref{16}) into (\ref{10}) and (\ref{12}), we get
\beq
\beta_G = \epsilon G - \frac{2(24-N)}{3} G^2. \label{17}
\eeq
Taking into account that the non-diagonal terms $G_{\varphi\sigma}$
should not be considered in (\ref{12}), and choosing the particular
{\it Ansatz} $\phi (\varphi)=\lambda\varphi$, one finds (by
substituting (\ref{16}) and (\ref{15}) into (\ref{12}))
\beq
\beta_{G_{\chi\chi}} = G(1- e^{2\varphi}) \left[ \frac{24-n}{3} \,
2\lambda +
\frac{N\epsilon}{2(\epsilon +1)} \lambda (\lambda -1) (\lambda-2)
\right].  \label{18}
\eeq
As the one-loop effective action (\ref{2}) is different from the
result one obtains in covariant gauges, Eq. (\ref{18}) differs
slightly from the corresponding results in Refs. \cite{5,6}.

Solving Eq. (\ref{17}) we find that $G=0$ is an IR stable fixed
point and that
\beq
G^* = \frac{3\epsilon}{2(24-n)}, \qquad \epsilon >0, \qquad N < 24,
\label{19}
\eeq
is a non-trivial ultraviolet fixed point. Both $\beta_G$ and the
corresponding fixed point $G^*$ are gauge parametrization
independent, as they should be.

 From (\ref{18}) we find that, due to the stringy form of
$\Gamma_{div}$, the non-trivial perturbative fixed point of order
$\lambda^* \sim \epsilon$ found in Ref. \cite{5} (it had been
pointed out already \cite{6} that the position of this fixed point
depends very much on the gauge) disappears completely, and only the
solution $\lambda^*=0$ remains. However, the other imaginary
ultraviolet fixed points, of order $\lambda^* \sim \epsilon^{-1/2}$,
 mentioned
in Ref. \cite{5} show up here as well:
\beq
\lambda_{1,2} = \pm 2i \sqrt{\frac{24 -N}{3N \epsilon}} + {\cal O}
(\epsilon^0).
\label{20}
\eeq
We have thus shown that in the string-like formalism the non-trivial
ultraviolet fixed point $\lambda^* \sim \epsilon$ of
dilatonic gravity (which was observed in the other formalism) does
not appear in the present model, which yields simply the `trivial'
solution $\lambda^* =0$.

\section{Discussion}

Probably the main result that can be extracted from the preceding
 study of a special type of sigma model near two dimensions is the
one that comes about from the consideration of the specific example of
dilatonic gravity in $d=2+\epsilon$ using the stringy
parametrization. Namely, the fact that the position of the UV fixed point
for the scalar dilaton coupling has collapsed to zero. In this way we
see that the non-trivial fixed point corresponding to this coupling
constant ---which had been found in previous works by different
authors (in other parametrizations)--- turns out to be, in the end, a
trivial fixed point. At the same time, the position of the fixed
point corresponding to the gravitational coupling constant does not
change. In any case, the {\it nature} of the $\lambda$ fixed point
does not change either ---it continues to be a UV saddle point as in Refs.
\cite{5}-\cite{9}.

It would be of interest to develop a RG formalism for the study
of more general sigma models than the class of sigma models
considered in the present work ---near two dimensions and in a
unified stringy way. Then one could hope to understand which type
of fixed point for the scalar-dilaton coupling (the trivial or the
non-trivial one) turns out to be the most acceptable physically.
In particular, adopting the point of view that string theory may prove to
be the fundamental `theory of everything', we believe that the trivial
fixed point for the scalar-dilaton coupling is in fact the physical UV
fixed point.

 \vspace{5mm}


\noindent{\large \bf Acknowledgments}

We  would like to thank A. Chamseddine for valuable discussions.
This work has been supported by DGICYT (Spain), project Nos.
PB93-0035
and SAB93-0024, by CIRIT (Ge\-ne\-ra\-litat de Catalunya), and by ISF
(Russia), grant RI1000.

\begin{appendix}

\renewcommand{\theequation}{{\mbox A}.\arabic{equation}}

\section{Appendix}

\setcounter{equation}{0}

In this Appendix we will show that the one-loop off-shell effective
action  ---which depends both on the gauge and on the
parametrization--- in dilatonic gravity (\ref{13}) leads to the
same on-shell result. As it follows form (\ref{16}), in the stringy
parametrization of dilatonic gravity the one-loop effective action
is given by
\beq
\Gamma_{div} = \frac{1}{2\pi \epsilon} \int d^2 x\,  \sqrt{-g}
\left\{{24-N \over 12} R + \frac{N}{2} \left[ (\phi')^2 (\varphi) -
 \phi'' (\varphi) - 2\phi'(\varphi) \right]
g^{\mu\nu} \partial_\mu \varphi \partial_\nu \varphi  \right\}.
\label{a1}
\eeq
At the same time, in the covariant gauge of Refs. \cite{8,6}, the
result of the calculation of $\Gamma_{div}$ in the same theory is
\beq
\Gamma_{div} = \frac{1}{2\pi \epsilon} \int d^2 x\,  \sqrt{-g}
\left\{{24-N \over 12} R - \frac{1}{2} \left[ 8- N (\phi')^2
(\varphi)  \right]
g^{\mu\nu} \partial_\mu \varphi \partial_\nu \varphi  \right\}.
\label{a2}
\eeq
As we clearly see, these two off-shell effective actions differ
because of their gauge and parame\-triz\-ation dependences.

Working, for simplicity, in the conformal gauge ($\bar{g}_{\mu\nu}
=\eta_{\mu\nu}$), we use with (\ref{a1}) the classical equations of
motion:
\bea
\frac{\delta L}{\delta \varphi} &=& \frac{1}{16\pi G} e^{-2\varphi}
(-2\Delta \sigma) - \frac{1}{2} \phi'  e^{-2\phi} (\partial
\chi_a)^2 =0, \nn \\
\frac{\delta L}{\delta \sigma} &=& \Delta \,  e^{-2\varphi} =0.
\label{a3}
\eea
 From the second of Eqs. (\ref{a3}) we find that the term
$( \phi'' + 2\phi')g^{\mu\nu} \partial_\mu \varphi \partial_\nu
\varphi$ in (\ref{a1}) is a boundary term and can therefore be
dropped out. Moreover, the term $-4 (\partial \varphi)^2$ in
(\ref{a2}) is also a boundary term. As a result, all (\ref{a1}),
(\ref{a2}) and the corresponding $\Gamma_{div}$ of Ref. \cite{5}
lead to a unique result on-shell (after dropping boundary terms),
which is
\beq
\Gamma_{div}^{on-shell} = \frac{1}{2\pi \epsilon} \int d^2 x\,
\sqrt{-g} \, \frac{N}{2}  (\phi')^2 (\varphi)
(\partial_\mu\varphi)^2.
 \label{a4}
\eeq
Hence, the corresponding $S$-matrix is in fact both gauge and
parametrization independent, as it should be from general considerations.

\end{appendix}


\newpage

\begin{thebibliography}{99}

\bibitem{4} S. Weinberg, in {\it General Relativity, an Einstein
        Centenary Survey}, eds. S.W. Hawking and W. Israel
        (Cambridge Univ. Press, Cambridge, 1979);
 R. Gastmans, R. Kallosh and C. Truffin, Nucl. Phys. {\bf B133}
        (1978) 417; S.M. Christensen and M.J. Duff, Phys. Lett.
{\bf B79} (1978) 213; I. Jack and D.R.T. Jones, Nucl. Phys. {\bf
B358} (1991) 695;  J. Nishimura, S. Tamura and A. Tsuchiya,
Mod. Phys. Lett. {\bf A9} (1994) 3565; T. Aida, Y. Kitazawa, H.
Kawai and M. Ninomiya, Nucl. Phys. {\bf B427} (1994)  158.

\bibitem{5} S. Kojima, N. Sakai and Y. Tanii, Nucl. Phys. {\bf
B246} (1994) 223.

\bibitem{6} E. Elizalde and S.D. Odintsov, preprint hep-th/9501015;
Phys. Lett. {\bf B347} (1995) 211.

\bibitem{9} Y. Tanii, S. Kojima and N. Sakai, Phys. Lett. {\bf
B322} (1994) 59; hep-th/9505132.

\bibitem{1}C. Callan, D. Friedan, E.J. Martinec and M. Perry, Nucl.
Phys. {\bf B262} (1985) 593.

\bibitem{2} E.S. Fradkin and A.A. Tseytlin, Nucl. Phys. {\bf B261}
(1985) 1.

\bibitem{10} L. Alvarez-Gaum\'e, D.Z. Freedman and S. Mukhi,
Ann. Phys.  {\bf 134} (1981) 85.

\bibitem{11} D. Friedan, Ann. Phys.  {\bf 163} (1985) 318.

\bibitem{3}  A.A. Tseytlin, Nucl. Phys. {\bf B294} (1987) 383.

\bibitem{7} A.H. Chamseddine, Nucl. Phys. {\bf B368} (1992) 98.

\bibitem{8} S.D. Odintsov and I.L. Shapiro, Phys. Lett. {\bf B263}
(1991) 183; Int. J. Mod. Phys. {\bf D1} (1993) 571.

\end{thebibliography}
\end{document}